\documentclass[
    aps, prb,
    reprint,
    english,
    superscriptaddress,
    longbibliography,
    floatfix,
    showkeys,                       
]{revtex4-1}

\usepackage[utf8]{inputenc}
\usepackage[T1]{fontenc}

\usepackage[%
]{graphicx}
\graphicspath{{Figs/}}
\usepackage{hyperref}

\usepackage{
    amsmath,
    amssymb,
    gensymb,
}

\begin{document}

\title{Thermal simulation of magnetization reversals for size-distributed assemblies of core-shell exchange biased nanoparticles}

\author{J. Richy}
\affiliation{
    Laboratoire de Magnétisme de Bretagne, Department of Physics, CNRS-UBO, 29285 Brest Cedex 3, France
}
\affiliation{
    Department of Physics, University of Johannesburg, PO Box 524, Auckland Park, Johannesburg 2006, South Africa
}
\author{J. Ph. Jay}
\author{S. P. Pogossian}
\author{J. Ben~Youssef}
\affiliation{
    Laboratoire de Magnétisme de Bretagne, Department of Physics, CNRS-UBO, 29285 Brest Cedex 3, France
}
\author{C. J. Sheppard}
\author{A. R. E. Prinsloo}
\affiliation{
    Department of Physics, University of Johannesburg, PO Box 524, Auckland Park, Johannesburg 2006, South Africa
}
\author{D. Spenato}
\author{D. T. Dekadjevi}
\email[]{david.dekadjevi@univ-brest.fr}
\affiliation{
    Laboratoire de Magnétisme de Bretagne, Department of Physics, CNRS-UBO, 29285 Brest Cedex 3, France
}

\date{\today}

\begin{abstract}
    A temperature dependent coherent magnetization reversal model is proposed for size-distributed assemblies of ferromagnetic nanoparticles and ferromagnetic-antiferromagnetic core-shell nanoparticles.
    The nanoparticles are assumed to be of uniaxial anisotropy and all aligned along their easy axis.
    The thermal dependence is included by considering thermal fluctuations, implemented via the Néel-Arrhenius theory.
    Thermal and angular dependence of magnetization reversal loops, coercive field and exchange-bias field are obtained, showing that F-AF size-distributed exchange-coupled nanoparticles exhibit temperature-dependent asymmetric magnetization reversal.
    Also, non-monotonic evolutions of \He and \Hc with $T$ are demonstrated.
    The angular dependence of \Hc with $T$ exhibits a complex behavior, with the presence of an apex, whose position and amplitude are strongly $T$ dependent.
    The angular dependence of \He with $T$ exhibits complex behaviors, which depends on the AF anisotropy and exchange coupling.
    The resulting angular behavior demonstrates the key role of the size distribution and temperature in the magnetic response of nanoparticles.

\end{abstract}


\maketitle

\section{Introduction}
    Over recent decades, magnetic nanoparticles (MNPs) have attracted a great deal of attention.\cite{Dormann-1997-ID516,  Dormann-1992-ID519}
    Indeed, these nanostructures exhibit different properties compared to bulk materials.
    This is firstly due to their reduced dimensions, secondly through the presence of interfaces, and thirdly via the inter-particle interactions.\cite{Petracic-2010-ID491}
    Extensive research has been performed on the fundamental aspects of these properties and on technological applications such as magnetic recording and magnetic hyperthermia.
    The last mentioned is the field of nanomedicine treating cancer by supplying heat to tumor cells using magnetic nanoparticles and an alternating magnetic field.\cite{Obaidat-2015-ID419, Gilchrist-1957-ID445, Kumar-2011-ID446, Issa-2013-ID517, Banobre-Lopez-2013-ID521, Dutz-2014-ID520}
    One of the key points concerning the applications of MNPs is their magnetization reversal thermal dependence.
    In the field of magnetic hyperthermia, the heating efficiency of a particle is represented by the specific absorption rate (SAR) which is directly related to the temperature $T$ and frequency-dependent hysteresis loop area (through hysteresis losses).\cite{Obaidat-2015-ID419, Carrey-2011-ID217}
    In the field of magnetic recording, a modification of the magnetization reversal over a small temperature range has key relevance for issues in applications.
    Firstly, such a change could be inferred by laser heating or applied current.
    Secondly, a device could be compromised by temperature fluctuations in a magnetic field.
    In addition, superparamagnetism is to be avoided in magnetic recording as it causes thermal destabilization of the recording units.\cite{Weller-1999-ID488}

    Consequently, at present there is interest of understanding and determining temperature-dependent magnetization reversals for a realistic assembly of magnetic nanoparticles.
    In order to obtain this understanding, a particle size distribution should be incorporated in a realistic model.
    Indeed, previous experimental studies have shown that there will be some size distributions of the MNPs regardless of the synthesis method used.\cite{Hergt-2006-ID447}
    This is still the case despite the fact that controlled synthesis has vastly improved in recent decades, together with advanced characterization methods.
    In the field of magnetic hyperthermia, the size dispersion drives the SAR.\cite{Vallejo-Fernandez-2013-ID420}
    The size distribution of MNPs was found to reduce the heating efficiency.\cite{Rosensweig-2002-ID421}
    In the field of magnetic recording, the size distribution was found to alter the uniformity of magnetic properties and consequently compromises the accompanying technologies associated with hard disks and proposed components of patterned media.\cite{Zhou-2000-ID518}

    The magnetization reversals of both a size-dispersed assembly of MNPs consisting of a single ferromagnetic core, as well as that of F-AF core-shell nanoparticles are of interest for applications.
    Indeed, the F-AF core-shell nanoparticles can exhibit a magnetic interfacial exchange coupling between the core and the shell, resulting in modification of the magnetization reversal and its temperature dependence.\cite{Evans-2009-ID352}
    In the field of magnetic hyperthermia, the SAR of the F-AF core-shell structures were found to be nearly one order of magnitude larger than those of MNPs of the core or shell materials alone.\cite{Hergt-2008-ID424, Mehdaoui-2011-ID422}
    It is suggested that the exchange coupling at the interface can be tuned in F-AF particles with the goal of enlarging the hysteresis area for an enhanced SAR.\cite{Lee-2011-ID489}
    In the field of magnetic recording,\cite{Moser-2002-ID490} F-AF core-shell nanoparticles are promising for enhancing the thermal stability of magnetic bits, and overcoming the so-called \textit{superparamagnetic limit} of recording.\cite{Nogues-2006-ID413,Skumryev-2003-ID347}

    Although many F and F-AF core-shell structures have been investigated experimentally and theoretically, no model describing the magnetization reversal of a size-dispersed assembly of F and F-AF core-shell nanoparticles has yet been proposed.
    Indeed, despite the keyrole of size distributions, previous theoretical models have solely probed the thermal behavior of single sized F and F-AF nanoparticles.
    It can be mentioned here that previous models based on Monte-Carlo algorithm have probed the thermal behavior either of an assembly of interacting single sized nanoparticles,\cite{Du-2006-ID339} or of a single F or F-AF nanoparticle with complex intrinsic spin configurations.\cite{Iglesias-2005-ID414}
    Theoretical research works quantifying the temperature dependent magnetization reversal of an assembly of size dispersed magnetic nanoparticles are not reported until now.
    Consequently, no models were able to quantitatively predict and reproduce effects of the size distribution on the thermal behavior of the magnetization reversal whereas some of these effects are experimentally probed for many decades.
    It should also be noted that experimental studies on MNPs often assume that the size distribution is at the origin of some thermal dependent magnetization reversal phenomena such as the thermal dependence of critical fields, whereas interparticle interactions could also lead to similar phenomena.
    Providing a model including size dispersion would contribute to solve such a controversy for a given assembly of mnps, as the contribution due to size dispersion could then be quantified.

    In this manuscript, a theoretical model and study of the magnetization reversal temperature dependence for size-distributed F and F-AF core-shell particles is presented.
    It includes the $T$ dependence of the reversal loops, as well as their angular dependence on the direction of the applied magnetic field (H).
    It should be noted here that one of the largest possible differences between experimental situation and a model is the controlled or uncontrolled misalignment between a given particle axis and the applied magnetic field.
    Probing the angular dependence of the reversal loops would include this into consideration.

    The first section of this paper describes the theoretical model.
    In the second section, the temperature dependence of ferromagnetic-only particles is studied, including size distribution.
    In the third section, the temperature dependence of F-AF nanoparticle magnetization reversal is studied, including an interfacial exchange coupling and a size dispersion.
    The results presented in section 2 and section 3 reveals some thermal dependent phenomena not yet predicted, such as the presence of apexes and reduced coercive fields greater than one.

\section{Thermally activated exchange bias model}

    In this section, a thermally-dependent exchange bias model for a single F-AF core-shell nanoparticle is described.

\begin{figure}
        \includegraphics{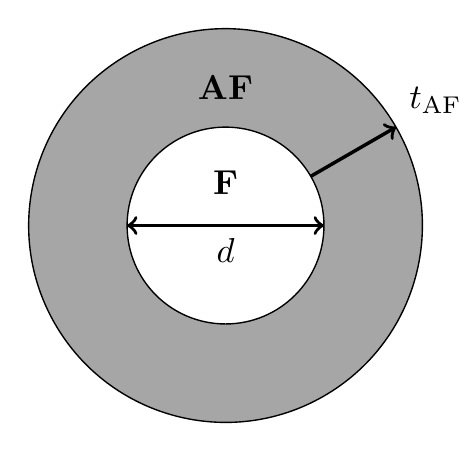}
        \caption{
            \label{geometry}
            Geometry of the F-AF core-shell particle, with $d$ the F core diameter and $t_{\mathrm{AF}}$ the AF shell thickness.
        }
    \end{figure}

\begin{figure}
        \includegraphics{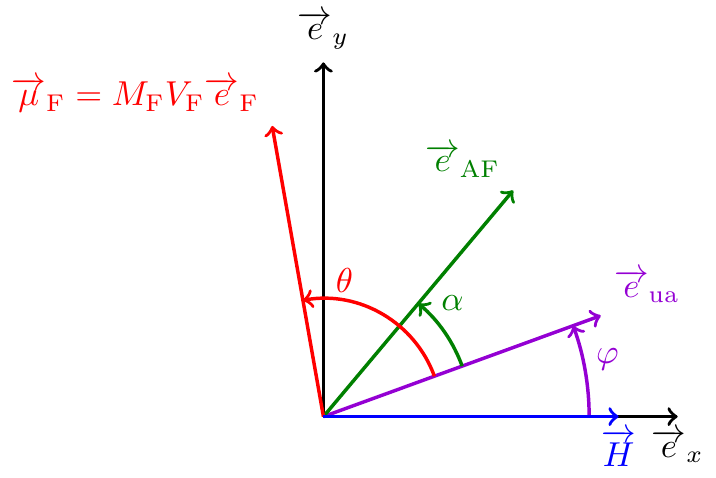}
        \caption{
            \label{coord_system}
            Diagram of the system axes, with $\roarrow e_\mathrm{F}$, $\roarrow e_\mathrm{AF}$, $\roarrow e_\mathrm{ua}$ and $\roarrow H$ the F domain, AF domain, anisotropy and field axes respectively.
            The field direction $\roarrow H$ also defines the measurement direction of the magnetic moment.
        }
    \end{figure}

    In the following, the F core diameter and AF shell thickness are represented by $d$ and \taf respectively, as shown in Fig.~\ref{geometry}.
    The interface between F and AF volumes ($V_\mathrm{F}$ and $V_\mathrm{AF}$ respectively) is a spherical surface \Sfaf.
    All vectors and angles involved are illustrated in Fig.~\ref{coord_system}, for a two-dimensional coordinate system.
    The F domain has a magnetic moment $\roarrow \mu_\mathrm{F} = M_\mathrm{F} V_\mathrm{F} \, \roarrow e_\mathrm{F}$, with an angle $\theta$.
    The AF domain has no net magnetization, but the antiparallel spins sublattice defines a direction $\roarrow e_\mathrm{AF}$ and can rotate with an angle $\alpha$.
    Both F and AF domains possess an uniaxial anisotropy axis $\roarrow e_\mathrm{ua}$, which, for reasons of simplicity, are considered colinear.
    The anisotropy axis is located at an angle $\varphi$ from the external field $H \roarrow e_\mathrm{x}$.

    The total free energy per particle $\mathcal{F}$ is then given by:
    \begin{align}
    	\begin{split}
    		\mathcal{F}(\theta, \alpha)  = {} & -\mu_0 H \times M_\mathrm{F} V_\mathrm{F} \,\cos(\theta + \varphi)\\
    		& + K_\mathrm{F} V_\mathrm{F} \; \sin^2\theta\\
    		& + K_\mathrm{AF} V_\mathrm{AF} \; \sin^2 \alpha\\
    		& - J_\mathrm{eb} S_\mathrm{F-AF} \; \cos(\theta - \alpha).\\
    	\end{split}
        \label{freeE}
    \end{align}

    Here, $K_\mathrm{F}$ and $K_\mathrm{AF}$ are the F and AF anisotropy constants respectively,
    and $J_\mathrm{eb}$ is the surface exchange energy.

    Equation~\ref{freeE} is based on the Meiklejohn and Bean\cite{Meiklejohn-1956-ID35, *Meiklejohn-1957-ID34} (MB) model, where the magnetization reversal is coherent for both F and AF domains.

    All magnetic parameters ($M_\mathrm{F}$, $K_\mathrm{F}$, $K_\mathrm{AF}$ and $J_\mathrm{eb}$) are considered as temperature independent.
    It should be noted that previous studies\cite{Jonsson-1997-ID402, Craig-2008-ID241}
    have shown that such an approximation is reasonable if no structural or magnetic transitions appear in the probed temperature range, and if the temperatures probed are not in proximity of the F and AF ordering temperatures (Curie or Néel temperature).

    The physical parameters associated with the F and AF core-shell particles are listed in Tab.~\ref{tab:parameters}. These are used for all further numerical calculations in this paper.
    It should be noted that F and AF physical parameters may vary largely from one system to another, especially when reducing experimental materials to the nanoscale.
    Even so, $M_\mathrm{F}$ and $K_\mathrm{F}$ in Tab.~\ref{tab:parameters} correspond to reported values of Fe.\cite{Kittel-2005-ID433, Sort-2004-ID353}
    The $K_\mathrm{AF}$ constant is in good agreement with AF oxides.\cite{Goya-2004-ID431, Ruette-2004-ID492, Vallejo-Fernandez-2007-ID416}
    The surface exchange energy $J_\mathrm{eb}$ is comparable to previous reported studies where it was found to vary largely between \SIrange{3}{840}{\micro J / m^2} for oxide based AF.\cite{Nogues-1999-ID92}
    Finally, previous studies have reported AF shell thicknesses below a few nanometers.\cite{Skumryev-2003-ID347, Iglesias-2007-ID399}
    Thus, these parameters are representative of actual nanoparticle systems.

    Additionally, to achieve a compromise between accuracy and speed, the simulation resolution is
    \SI{0.1}{\degree} in $\theta$ for F-only particles,
    and \SI{0.5}{\degree} and \SI{2}{\degree} in $\theta$ and $\alpha$ for F/AF particles respectively.
    A magnetic field resolution of \SI{0.2}{mT} was used for both calculations.

    \begin{table}[t]
        \caption{
            \label{tab:parameters}
            Physical parameters used for the F-only and F-AF core-shell particles.
            A second set of parameters is indicated by the $^*$ exponent.
        }
        \begin{ruledtabular}
            \begin{tabular}{cccc}
                \multicolumn{2}{c}{\bfseries F parameters}                    & \multicolumn{2}{c}{\bfseries AF parameters}\\
                \colrule
                $<d>$   & \SI{8}{nm}        & \taf       & \SI{0.5}{nm}             \\
                \Mf     & \SI{1740}{kA/m}   & \Jeb       & \SI{75}{\micro J/m^2}    \\
                \Kf     & \SI{50}{kJ/m^3}   & \Kaf       & \SI{200}{kJ/m^3}         \\
                        &                   & $\Jeb^*$   & \SI{10}{\micro J/m^2}    \\
                        &                   & $\Kaf^*$   & \SI{100}{kJ/m^3}         \\
            \end{tabular}
        \end{ruledtabular}
    \end{table}

\begin{figure}
        \includegraphics{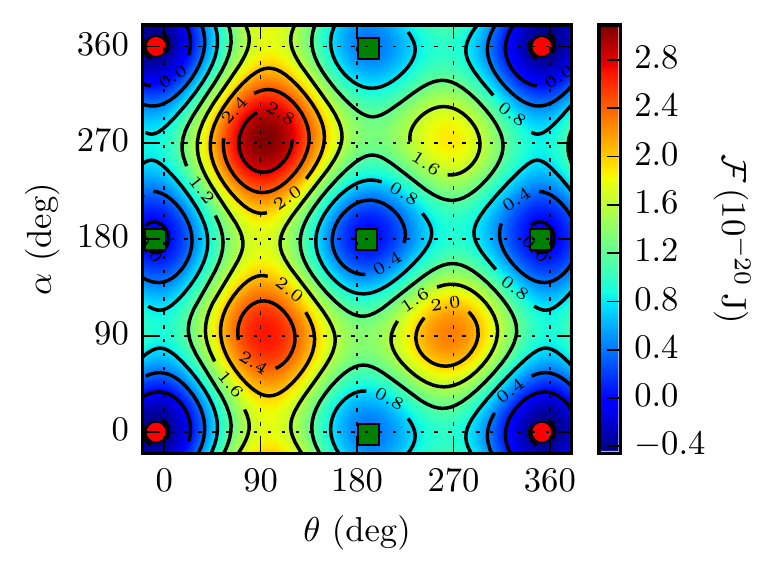}
        \caption{
            \label{2Dlandscape}
            Free energy landscape $\mathcal{F}$, with $\mu_0 H = \SI{15}{mT}$ and $\varphi = \SI{60}{\degree}$.
            \rond, stable equilibrium; \carrevert, metastable equilibrium.
            \trait{black} represent $\mathcal{F}$ iso-energy contour lines.
        }
    \end{figure}

    The analysis of $\mathcal{F}$ shows that for a given value and direction of the magnetic field $(H, \varphi)$,
    the energy landscape is composed of one or multiple local minima, separated by saddle-points.
    All the particle states $(\alpha, \theta)$ around the minimum and below the lowest surrounding saddle-point energy level define a valley region, where the F-AF directions can be trapped.
    Figure~\ref{2Dlandscape} shows an example of the free energy landscape, with four different valley regions around the red and green markers identifying local equilibria.
    By changing the field magnitude, the energy landscape evolves, modifying the different valley minima and also their corresponding energy barriers.
    In particular, at high field (i.e. above saturation), only one valley remains.

    From one valley to another, when varying H, the particle equilibrium energy state has to get over the corresponding saddle-point energy barrier.
    Following this method, an equilibrium ($\alpha, \theta$ couple) can be determined for each field value, and its evolution defines the magnetization reversal.

    In order to introduce a thermal agitation, a transition from one valley to another is determined using the Néel-Brown relaxation theory.\cite{Neel-1949-ID356, Coffey-2012-ID381}
    Knowing the energy barrier $\Delta E$ of a local minimum of a valley, the time-dependent probability of remaining in this valley is proportional to $\exp(-t / \tau)$ with $t$ and $\tau$ the time and relaxation time, respectively.
    $\tau$ is given by Néel's law:
    \begin{equation}
        \tau = \frac{1}{f_0} \exp\left(\frac{\Delta E}{\kb T}\right)
        \label{neel},
    \end{equation}

    with $f_0$ the attempt frequency (of the order of \SI{e10}{Hz}),\cite{Coffey-2012-ID381} $\kb$ the Boltzmann constant, and $T$ the temperature.

    Introducing a characteristic measurement time $\tau_{\mathrm{meas}}$, it is commonly accepted that the transition to a new valley occurs when the relaxation time $\tau$ is below or of the same order of magnitude as $\tau_{\mathrm{meas}}$, which leads to
    \begin{equation}
        \Delta E < \ln(f_0 \tau_{\mathrm{meas}}) \times \kb T
        \label{Ineq}.
    \end{equation}
    This criterion is generally approximated to $\Delta E < 25 \kb T$.\cite{Bean-1959-ID314, Nunes-2004-ID263}

    The method for choosing the equilibrium state considers the system state to be only in a single valley.
    Starting from a valley energy minimum and after a small field increment, the local minimum will have moved slightly to an energy level  $E_\mathrm{ini}$.
    As discussed previously, only an energy barrier lower than $25 \kb T$ compared to the reference valley can be overcome.
    The statistically accessible states domain $\mathcal{S}$ is defined by all the energy levels lower than $E_\mathrm{ini} + 25 \kb T$ and forming a connected domain including the starting valley.
    This accessibility domain may contain multiple valleys and their intermediate saddle-point energy barriers.

    Then the probability for the system to be in the $(\alpha, \theta)$ state is
    \begin{equation}
        P(\alpha, \theta) =
            \begin{cases}
                \displaystyle\frac{1}{Z} \exp\left(-\frac{\mathcal{F(\alpha, \theta)}}{\kb T} \right), & \text{if }(\alpha, \theta) \in \mathcal{S}\\
                0, & \text{otherwise.}
            \end{cases}
    \end{equation}

    Here $Z$ is the partition function defined by
    \begin{equation}
        Z = \displaystyle\int_\mathcal{S}  \exp\left(-\frac{\mathcal{F(\alpha, \theta)}}{\kb T} \right) \dd\theta \dd\alpha.
    \end{equation}

\begin{figure}
        \includegraphics{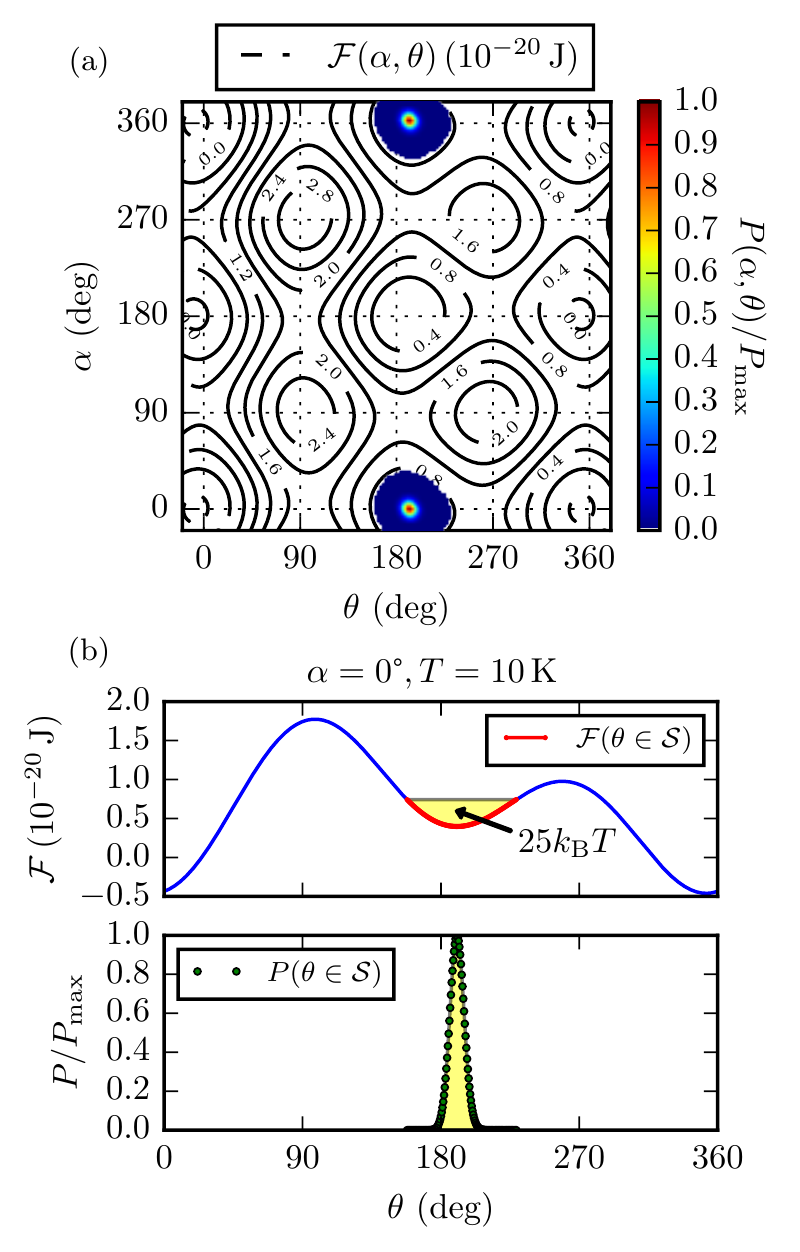}
        \caption{\label{probEnergy}
            (a) Normalized probability function (colored area) for $T = \SI{10}{K}$ with the energy landscape (black contour lines) from Fig. \ref{2Dlandscape}.
            (b) Slice for $\alpha = \SI{0}{\degree}$, with the energy function (upper) and the normalized probability (lower).\\
            Here, the current valley corresponds to a metastable position.
            The $\mathcal{S}$ domain is deduced by calculating all the energy states below $25 \kb T$ and including the current metastable equilibrium state.
        }
    \end{figure}

    Figure~\ref{probEnergy}(a) shows the example of the energy landscape from the previous figure at $T = \SI{10}{K}$.
    The accessible domain $\mathcal{S}$ is colored in, whereas all the remaining states with zero probability are shown without color.
    The colored sidebar gives the normalized probability value.
    At this temperature, there is only one accessible local minimum, as the thermal energy is not high enough to surpass the energy barriers.
    Figure~\ref{probEnergy}(b) corresponds to a slice for $\alpha = \SI{0}{\degree}$, with the upper part representing the energy level and the lower part showing the normalized probability.

    The most probable state is then the lowest valley of $\mathcal{S}$.
    If this valley is different from the starting one, the equilibrium state of the system changes and jumps to this new equilibrium.
    Otherwise, it remains in the current valley.
	This process avoids the need to determining all possible saddle-points, and is repeated as many times as necessary until the selected valley remains the same.

    \begin{figure*}
        \includegraphics{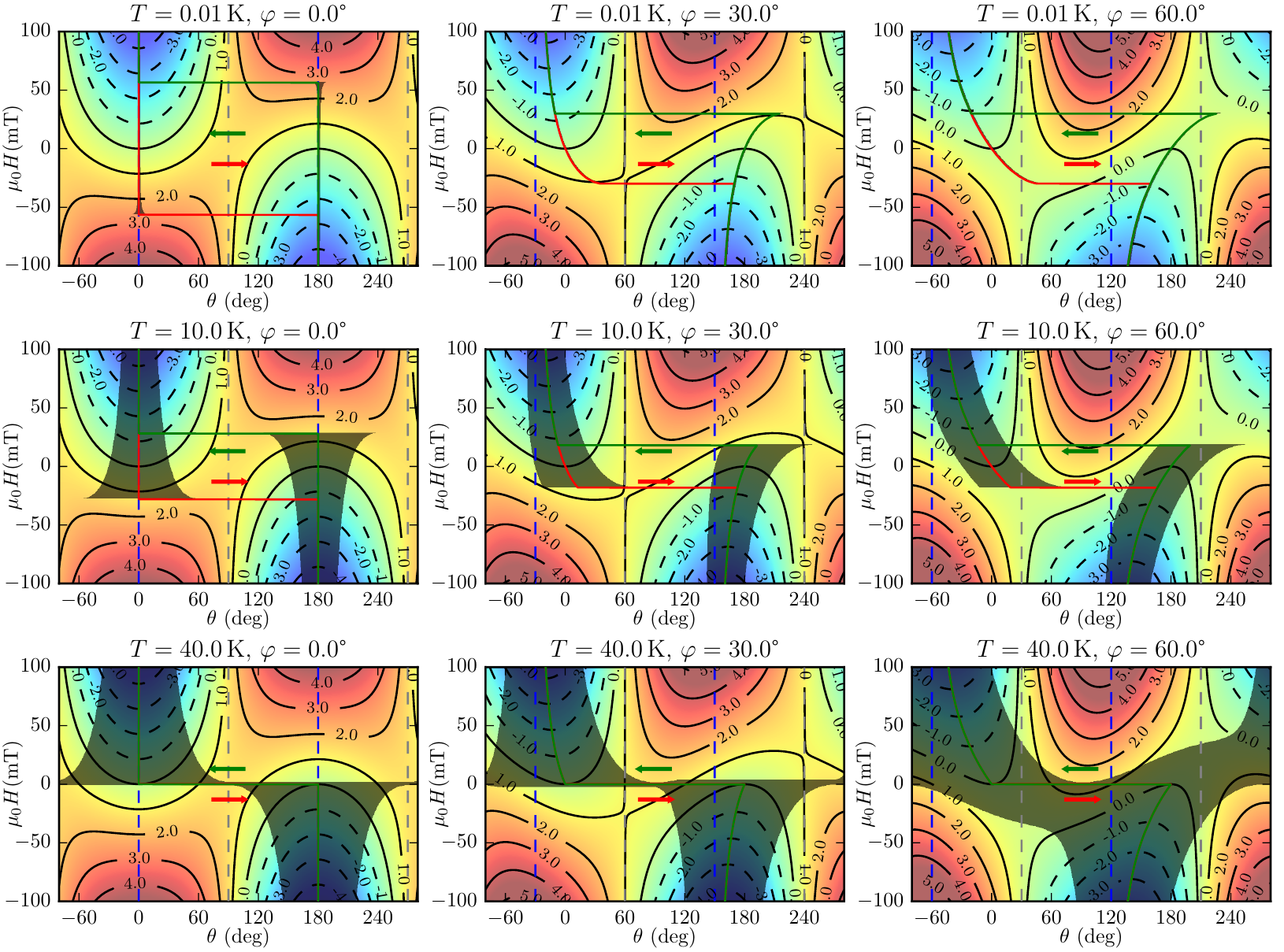}
        \caption{\label{MonoPart_energy}
            Evolution for a F particle of the magnetic moment direction $\theta$ with the applied field, for different field directions and at different temperatures.
            The \trait{red} and \trait{green} paths represent the decreasing and increasing branches, respectively.
            The colored map with iso-energy contour lines (units in \SI{e-20}{J}) corresponds to the energy landscape $\mathcal{F}(H, \theta)$, with all the accessible directions grayed out depending on the temperature.
            The blue and gray dashed lines are visual guides and represent the directions parallel and perpendicular to the field direction, respectively.
        }
    \end{figure*}

    Finally, the average magnetic moment $<\mu_\mathrm{F}>$ is deduced by integrating the projection of the magnetic moment on the field direction over all the accessible energy states $\mathcal{S}$:
    \begin{equation}
        <\mu_\mathrm{F}> = M_\mathrm{F} \int_\mathcal{S} P(\alpha, \theta) \cos(\theta + \varphi) \dd\theta\dd\alpha.
        \label{magmoment}
    \end{equation}

    Hence, bounds for the coercive fields, $H_\mathrm{c1}$ and $H_\mathrm{c2}$, can be found by examining the evolution of \muF with the external magnetic field.

    It should be noted that thermal agitation may be implemented in different ways.
    Indeed, two different approaches have been proposed:
    the magnetic moment can be considered either in a one-valley state approximation as presented above,\cite{Garcia-Otero-1998-ID218, Garcia-Otero-1999-ID300, Franco-2004-ID209}
    or in a multiple-valley state approximation (with transition rates between the different valleys).\cite{Pfeiffer-1990-ID429, Pfeiffer-1990-ID430, Iglesias-2002-ID428, Xu-2001-ID338, Du-2006-ID339, Carrey-2011-ID217}
    A multiple valley approximation is more difficult to process, as one needs to determine all the metastable positions and saddle-points between valleys, and resolve the time equation with all the energy levels as variable.
    Until now, this approach has only been used for a one-variable energy equation, whereas the free energy considered here in Eq.~\ref{freeE} includes two variables.

    For example, the free energy landscape in Fig.~\ref{2Dlandscape} has 4 accessible valleys and 8 different saddle-points, which requires the resolution of a differential equation with 12 independent variables.
    Therefore, a single-equilibrium approximation with local thermal fluctuation is used in this article, allowing the F-AF particle equilibrium state $(\alpha, \theta)$ to be in only one energy valley for a given $H,T$ couple.

\section{Ferromagnetic nanoparticles}

    This section presents the $T$ dependence of F-only particle magnetization reversal.
    Consequently, both F/AF exchange coupling \Jeb and AF anisotropy constant \Kaf are set to zero, the AF direction $\alpha$ is not considered, and the free energy function $\mathcal{F}$ only depends on $H$ and $\theta$.

    The $T$ dependence of \Hc of a single F nanoparticle is first presented so as to examine the results of the numerical approach described in the previous section.
    Following this, the $T$ evolution of the magnetization reversal angular dependence is discussed.
    Indeed, this angular dependence is a key property of the magnetization reversal as it provides details about axial or/and unidirectional anisotropic properties.
    Then, the temperature evolution of this angular dependence would provide key information on the axial and unidirectional anisotropic properties, so as to reveal the modifications of anisotropy configuration with temperature.\cite{Ambrose-1997-ID325, Kim-2000-ID486}

    Moreover, the presence of misalignments between anisotropy axes has been previously probed in exchange-biased systems using the MR (magnetization reversal) angular dependence.
    It provides key information on the ratio of anisotropy constants.\cite{Jimenez-2011-ID386, Zhang-2010-ID335, Dekadjevi-2011-ID97}
    A study on exchange-biased nanoparticles will be presented in section~\ref{sec:bias}.

    \subsection{Single FM particle}

        The energy landscape $\mathcal{F}(H, \theta)$ is shown in Fig.~\ref{MonoPart_energy} for an F-only particle, at different field directions and temperatures.

        At $T = \SI{0.01}{K}$ (in Fig.~\ref{MonoPart_energy}, first line), the magnetization reversal corresponds to the \citet{Stoner-1948-ID306} (SW) reversal cycle.
        Indeed, \muF follows the local valley and only experiences switching when the equilibrium disappears.
        For $\varphi = \SI{0}{\degree}$, i.e. with the field aligned with easy axis, the equilibrium position stays on the same axis, creating a squared hysteresis cycle.
        For non-zero $\varphi$, as represented in the figures shown in the second and third rows in Fig.~\ref{MonoPart_energy}, the equilibrium moves away from the field direction, reducing the measured $\muF \cos(\theta)$ and the $\mu_0 H$ at which \muF flips.

\begin{figure}
            \includegraphics{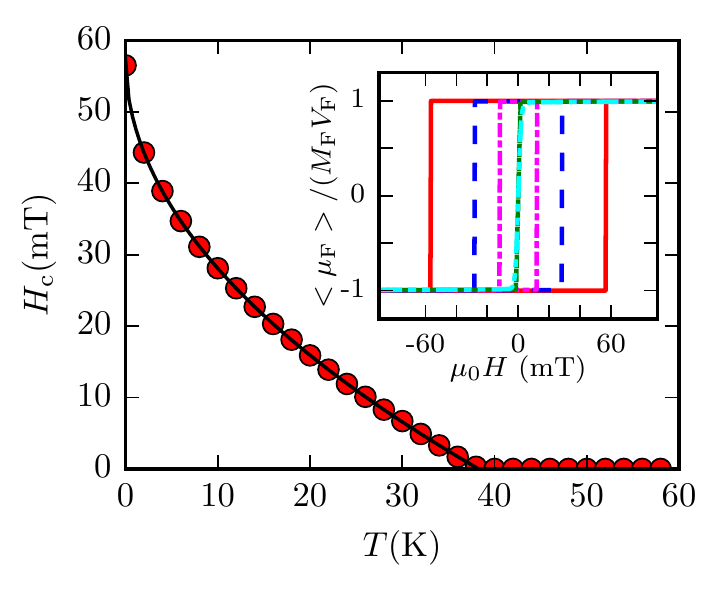}
            \caption{
                \label{cycle_single_part}
                \Hc temperature evolution (\rond) for $\varphi=\SI{0}{\degree}$ of a \SI{8}{nm} single F particle , along with the theoretical evolution (\trait{black}) using Eq.~\ref{eq:Hc_evol}.
                The inset shows corresponding reversal cycles for $T =$~\SI{0.01}{K}~(\trait{red}),  \SI{10}{K}~(\traitun{blue}), \SI{24}{K}~(\traitdeux{magenta}), \SI{40}{K}~(\traittrois{OliveGreen}) and \SI{58}{K}~(\traitquatre{Turquoise}).
            }
        \end{figure}

        At $T\neq \SI{0}{K}$, \muF can reach multiple states in the local valley, reducing the energy barrier by $25 \kb T$.
        Thus, \muF switches at a lower field strength, resulting in a decrease of \Hc with $T$, as shown in Fig~\ref{cycle_single_part}.
		As expected, the \Hc decrease obtained here follows the well known\cite{Nunes-2004-ID263, Garcia-Otero-1998-ID218} relationship:
		\begin{equation}
			\label{eq:Hc_evol}
			\Hc = \Hc^0 \left[ 1 - \sqrt{25 \kb T / (\Kf \Vf)} \right] ,
		\end{equation}

        where \Hc reaches zero when $\Kf \Vf = 25 \kb T$.
        Above this temperature defined as the blocking temperature, all the energy barriers are below $25 \kb T$ and the particles then have a superparamagnetic\cite{Petracic-2010-ID491} behavior.
        \muF can randomly flip direction under the influence of $T$. Thus, the time-average \muF is zero under zero field.
        Figure~\ref{MonoPart_energy} (third line) illustrates this behavior (i.e. magnetization reversal with no hysteresis).

\begin{figure}
            \includegraphics{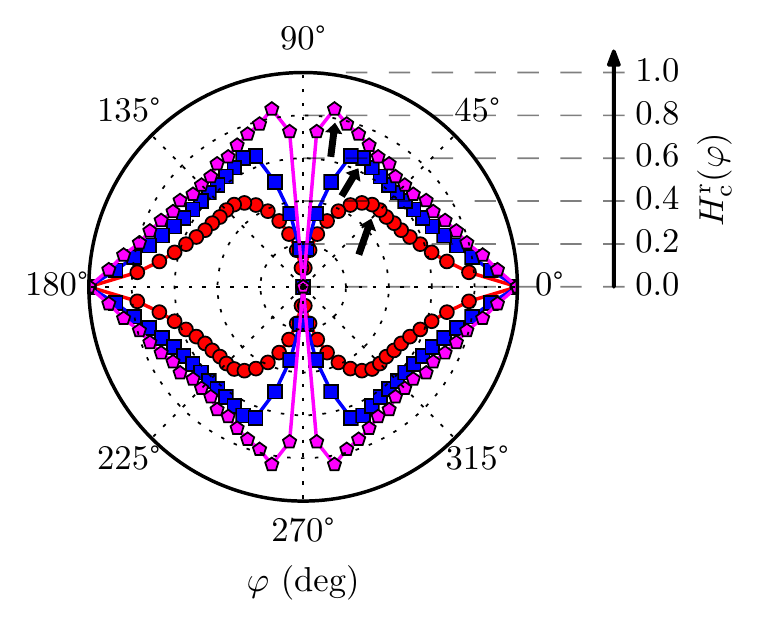}
            \caption{\label{MonoPart_HcEvol_azimuth}
                Azimuthal behavior of the reduced coercive field \Hcr for a single F particle at $T =$~\SI{0.01}{K}~(\rond), \SI{10}{K}~(\carre) and \SI{30}{K}~(\pentagon).
                The arrows identify the coercive apexes positions.
            }
        \end{figure}

        In the following, the angular dependence of the reduced coercive field \Hcr, defined as $H_\mathrm{c}^\mathrm{r}=\Hc(\varphi) / \Hc(\varphi = \SI{0}{\degree})$, is studied as a function of $T$ (see Fig.~\ref{MonoPart_HcEvol_azimuth}) for the \SI{8}{nm} single particle described in Tab.~\ref{tab:parameters}.
        At \SI{0.01}{K}, its angular dependence reveals a butterfly-like geometry as expected from the SW model.
        However, the increase of $T$ modifies the overall shape as \Hcr increases close to the hard axis.
        This results in a \textit{less rounded} angular shape of \Hcr.
        Also, there are coercive apexes defined as local \Hc maxima (arrows in Fig.~\ref{MonoPart_HcEvol_azimuth}).
        Their angular positions evolve with $T$ towards the hard axis, and their amplitude increase with $T$.
        These evolutions with $T$ arise from the difference between coercive bounds (field value at which $\muF$ is perpendicular to the field) and switching fields (which are defined as fields values at which the $\muF$ direction jumps\cite{Garcia-Otero-1998-ID218}).
        Indeed, at $\SI{0}{K}$, \Hc diverges from the switching field when $\varphi > \SI{45}{\degree}$, and falls gradually to zero, whereas the later increases to converge back to its highest value at $\varphi = \SI{90}{\degree}$.
        However, at a non-zero $T$, this divergence occurs at a higher angle, which explains the \Hc increase when $T$ approaches the blocking temperature.

        Whereas the SW model is often considered as a good approximation, it does not reproduce the evolutions discussed above.
        This non-zero temperature behavior should be considered in temperature dependence magnetization reversal studies.

    \subsection{Size-distributed FM particles}
    \label{subsec:distrib}

        In order to reproduce magnetic reversal behavior of actual systems, a large population of particles with different sizes should be included.
        Indeed, previous experimental studies have shown that the distribution of F particle sizes plays a key role for the understanding of ferromagnetic nanoparticle behavior.\cite{Jonsson-1997-ID402, Tiwari-2012-ID405, Khanna-2013-ID404, Pisane-2015-ID406}

        In the following, a lognormal distribution of particle diameters is used with a mean diameter $<\!d\!> = \SI{8.29}{nm}$ and standard deviation $\sigma_\mathrm{sd} =  \SI{2.99}{nm}$.
        It should be noted that previous studies have shown that lognormal distributions are often observed in experimental systems.\cite{Skumryev-2003-ID347, Sort-2004-ID353, Nishioka-1996-ID357, Pisane-2015-ID406, Tiwari-2012-ID405, Sort-2004-ID353}
        The nanoparticles are assumed to be all aligned along their easy axis.

\begin{figure}
            \centering
            \includegraphics{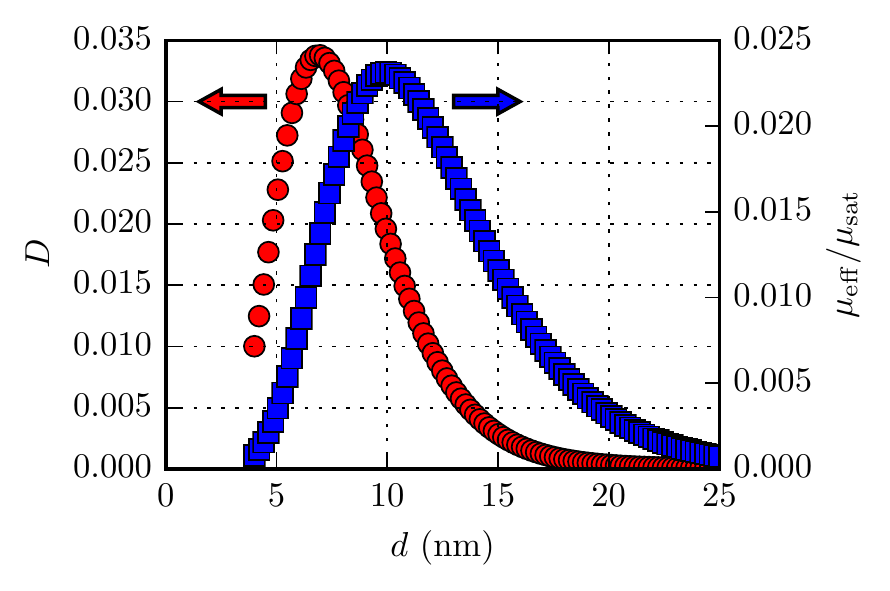}
            \caption{\label{domains_twinDM}
                Lognormal particle size distribution normalized with 100 discrete diameters, (\rond).
                Effective magnetic moment per particle for each corresponding diameter, normalized with the total magnetic moment $\mu_\mathrm{sat}$, (\carre).
            }
        \end{figure}

        Figure~\ref{domains_twinDM} (left axis, round red symbols) shows the normalized particle size distribution $D$ in the range of \SIrange{4}{25}{nm}, for 100 discrete $d$ values.
        Each diameter has a corresponding effective magnetic moment per particle $\mu_\mathrm{eff} = \mu_\mathrm{F} \times D$.
        The normalized effective moment $\mu_\mathrm{eff} / \mu_\mathrm{sat}$ is also shown in Fig.~\ref{domains_twinDM} (right axis), with $\mu_\mathrm{sat}$ defining the saturated moment of the particle assembly.
        It shows that diameters which drive the magnetic behavior of the whole particle population are not around the maximum of the particle distribution nor the particle mean size, but are a compromise between the number of particles at that diameter and the corresponding total magnetic moment.
        The key point here is the effective magnetic moment distribution shown in Fig.~\ref{domains_twinDM}.

        The process for calculating the reversal cycle of an assembly of size-distributed particles is then as follows:
        for each diameter among the 100 discrete values, a reversal cycle is calculated.
        Considering no inter-particle interaction, the total magnetic moment is determined by the summation of each particle's moment, weighted by the density distribution.

\begin{figure}
            \centering
            \includegraphics{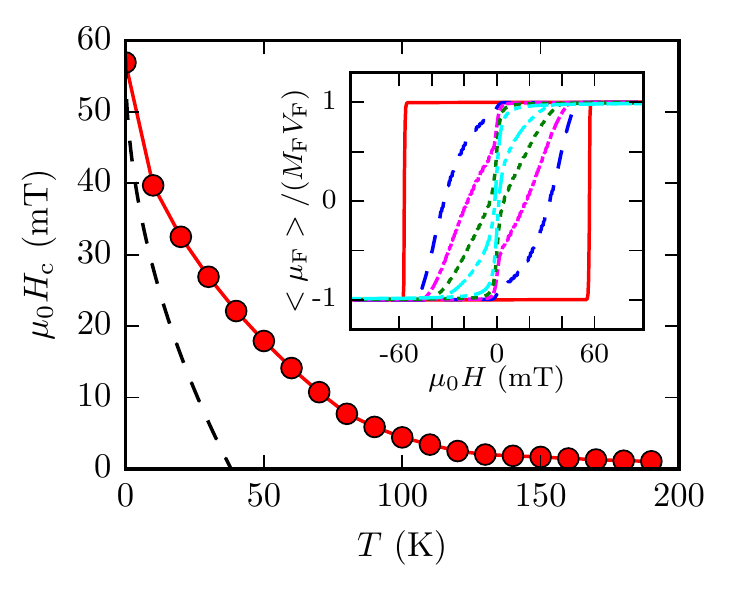}
            \caption{
                \label{Distrib_MFpart_phi0_evolT}
                \Hc temperature evolution (\rond) for a 100 discrete diameter size-distributed FM nanoparticle assembly, and \Hc evolution (\traitun{black}) of a \SI{8}{nm} single particle at $\varphi = \SI{0}{\degree}$ for comparison.
                The inset shows the corresponding reversal cycles for an assembly for $T = \SI{0.01}{K}$~(\trait{red}), \SI{20}{K}~(\traitun{blue}), \SI{50}{K}~(\traitdeux{magenta}), \SI{100}{K}~(\traittrois{OliveGreen}) and \SI{190}{K}~(\traitquatre{Turquoise}).
                $\varphi = \SI{0}{\degree}$ for all.
            }
        \end{figure}

        Figure~\ref{Distrib_MFpart_phi0_evolT} shows the obtained hysteresis cycles for multiple $T$ at $\varphi = \SI{0}{\degree}$.
        The critical temperature at which the coercivity is negligible for an assembly of particles (with a \SI{8}{nm} average diameter) is much greater than for a \SI{8}{nm} single diameter particle, as expected.
        The particle assembly exhibits a higher coercivity for a given $T$ due to the larger diameter population.

\begin{figure}
            \centering
            \includegraphics{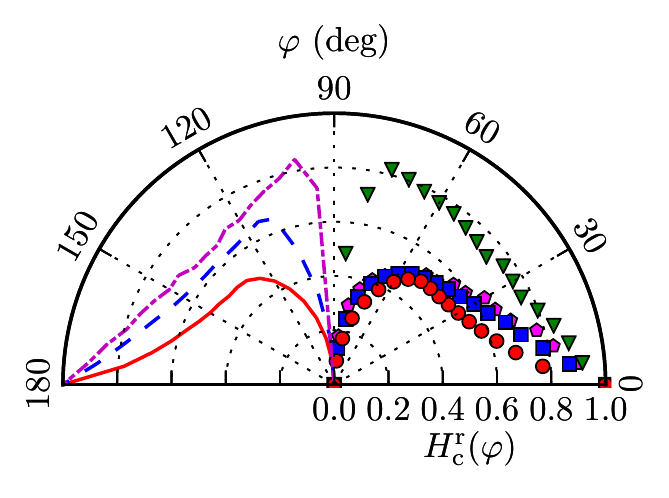}
            \caption{
                \label{Distrib_FMpart_azimuth}
                On the right quarter, azimuthal behavior of size-distributed particles for $T =$ \SI{0.01}{K} (\rond), \SI{20}{K} (\carre), \SI{50}{K}(\pentagon) and \SI{100}{K}(\trianglemark).
                On the left quarter, for comparison, are plotted \Hcr for a $\SI{8}{nm}$ single particle for $T =$~\SI{0.01}{K}~(\trait{red}), \SI{10}{K}~(\traitun{blue}) and \SI{30}{K}~(\traitdeux{magenta}).
            }
        \end{figure}

        The thermal azimuthal behavior of size-distributed particles in Fig.~\ref{Distrib_FMpart_azimuth} shows the same symmetry and shape as those obtained for a single particle in Fig.~\ref{MonoPart_HcEvol_azimuth}.
        Also, \Hcr are identical at \SI{0}{K}.
        It should be noted that the evolution of the apex observed for a single particle is still present for a particle assembly.

\section{Exchange biased nanoparticles}
    \label{sec:bias}

    This section focuses on core-shell exchange-coupled particles.
    The F core is now coupled to the AF shell with \Jeb, \Kaf and \taf values from Tab.~\ref{tab:parameters}.
    The fact that the AF domain is able to rotate is a driving mechanism for thermal dependence of magnetization reversal as explained below, and has only been simulated for thin film using a single domain\cite{Binek-2001-ID319, Lebeugle-2010-ID8} or multiple AF grains.\cite{Hou-2001-ID149}

    Indeed, in a conventional MB model, the AF layer is rigid and cannot rotate.
    The exchange energy creates unbalanced energy valleys in the same way as a field strength would.
    Thus, the reversal cycle is field-shifted along the field axis, of a value \He called the exchange-bias field.
    The thermal energy then only affects the switching field at which the reversal occurs.
    Hence, \He does not depend on $T$ and has a constant $\Jeb \Sfaf$ value for $\varphi = \SI{0}{\degree}$.

    However, when the AF spins are able to rotate, the exchange energy creates unbalanced energy valleys between the two stable directions of the AF spins as shown in Fig.~\ref{2Dlandscape}.
    Depending on the ratio between the \freeE energy terms, two different processes can be observed for a single F-AF particle.
    Below a temperature defined here as an exchange-blocking temperature where \He is not zero, the AF shell anisotropy barrier is higher than the thermal fluctuation.
    The F moment rotates without dragging the AF, and a shift of the hysteresis cycle is observed.
    At a higher temperature, when the AF anisotropy energy barrier is low enough for the thermal fluctuation to be dominant, both F and AF rotate simultaneously, and no field shift is observed.
    Instead, as the effective anisotropy includes both F and AF anisotropies, the coercive field increases.
    The temperature at which the two behavior transitions occur depends directly on the AF anisotropy energy and exchange energy.

\begin{figure}
        \centering
        \includegraphics{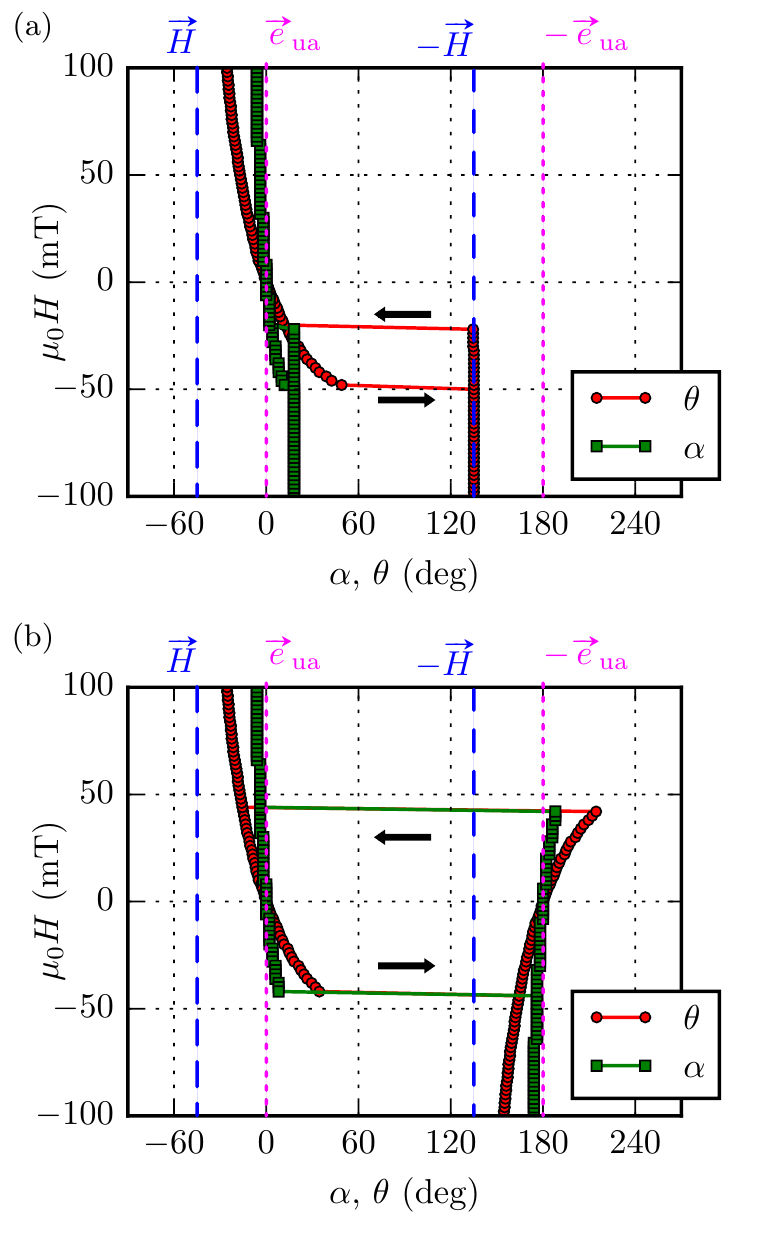}
        \caption{
            \label{fig:MonoPart_FMAM_2DPath}
            F and AF directions during a field cycle at $\varphi = \SI{45}{\degree}$, for (a) $T = \SI{0.01}{K}$ , and (b) $T = \SI{15}{K}$.
            Obtained \He values are \SI{-33.7}{mT} and \SI{0}{mT} respectively, and \Hc values are \SI{13.0}{mT} and \SI{42.3}{mT}.
            $\roarrow H$ (\traitun{blue}) and $\roarrow e_\mathrm{ua}$ (\traittrois{pymagenta}) represent the field and anisotropy axis directions, respectively.
        }
    \end{figure}

    The two possible reversal mechanisms are shown in Fig.~\ref{fig:MonoPart_FMAM_2DPath}(a) and \ref{fig:MonoPart_FMAM_2DPath}(b), representing the F and AF angle evolution with the applied field for the same parameters and at two different temperatures.
    For a high enough H, the F moment is dragged by H due to the Zeeman interaction, and the AF spin sublattice is dragged by the F moment due to the F-AF exchange interaction.
    At $\SI{0.01}{K}$, only the F moment flips, and a shift is observed.
    The AF direction $\alpha$ also experiences a small hysteretic evolution, as shown in Fig.~\ref{fig:MonoPart_FMAM_2DPath}(a) for $\mu_0 H \in [-50, -25] \si{mT}$.
    At $\SI{15}{K}$, all F and AF spins rotate simultaneously.
    The \Hc bounds are then symmetrical, and no \He exists.
    In-between these two different behaviors, \He and \Hc change abruptly due to the one-valley state approximation.

    \begin{figure}
        \centering
        \includegraphics{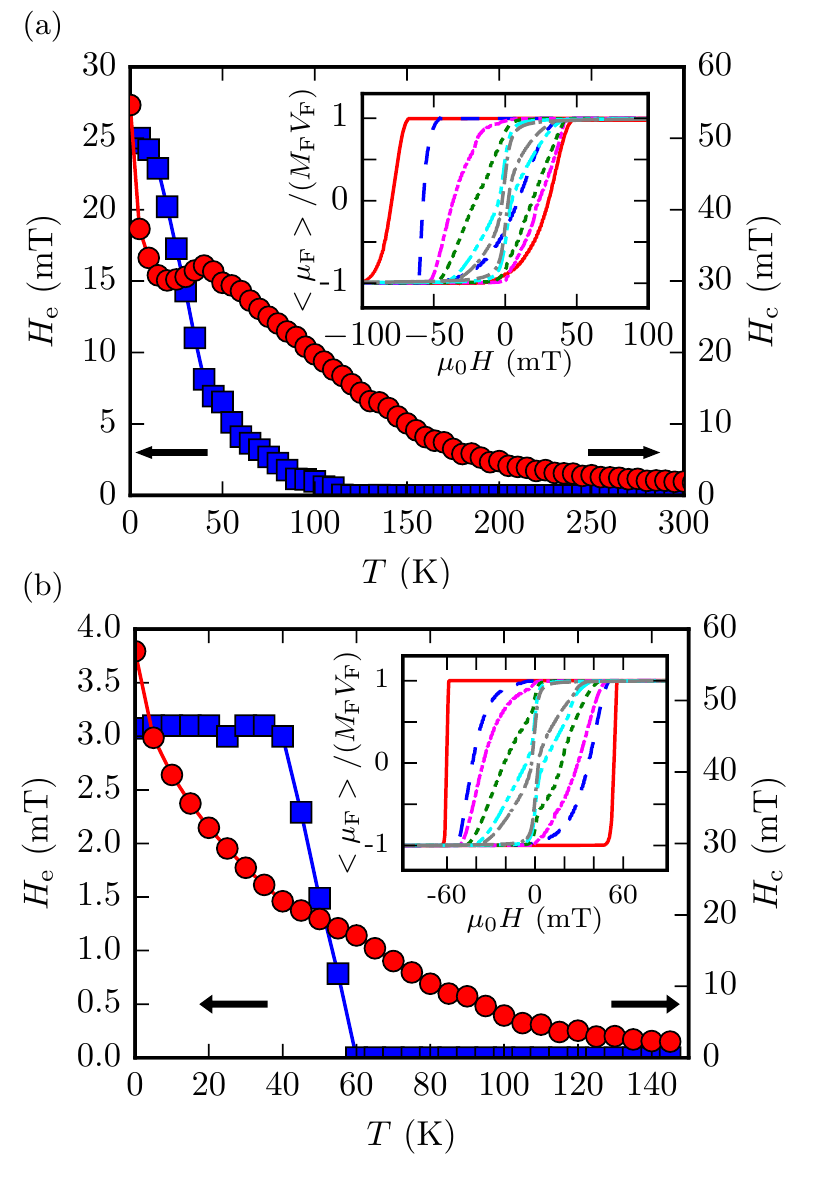}
        \caption{
            \label{fig:S_FMAF_Hevol}
            \Hc (red circles) and \He (blue squares) along the anisotropy axis for size-distributed F-AF core-shell particles.
            (a) corresponds to the (\Kaf, \Jeb) set of parameters.
            The inset shows corresponding reversal cycles for $T =$ \SI{0.01}{K}~(\trait{red}), \SI{10}{K}~(\traitun{blue}), \SI{50}{K}~(\traitdeux{magenta}), \SI{100}{K}~(\traittrois{OliveGreen}), \SI{200}{K}~(\traitquatre{Turquoise}) and \SI{300}{K}~(\traitcinq{gray}).
            (b) shows \Hc and \He evolution for the  ($\Kaf^*, \Jeb^*$) set, with the corresponding reversal cycles in the inset for $T =$ \SI{0.01}{K}~(\trait{red}), \SI{10}{K}~(\traitun{blue}), \SI{20}{K}~(\traitdeux{magenta}),
            \SI{50}{K}~(\traittrois{OliveGreen}), \SI{100}{K}~(\traitquatre{Turquoise}) and \SI{145}{K}~(\traitcinq{gray}).
        }
    \end{figure}

    \begin{figure*}
        \centering
        \includegraphics{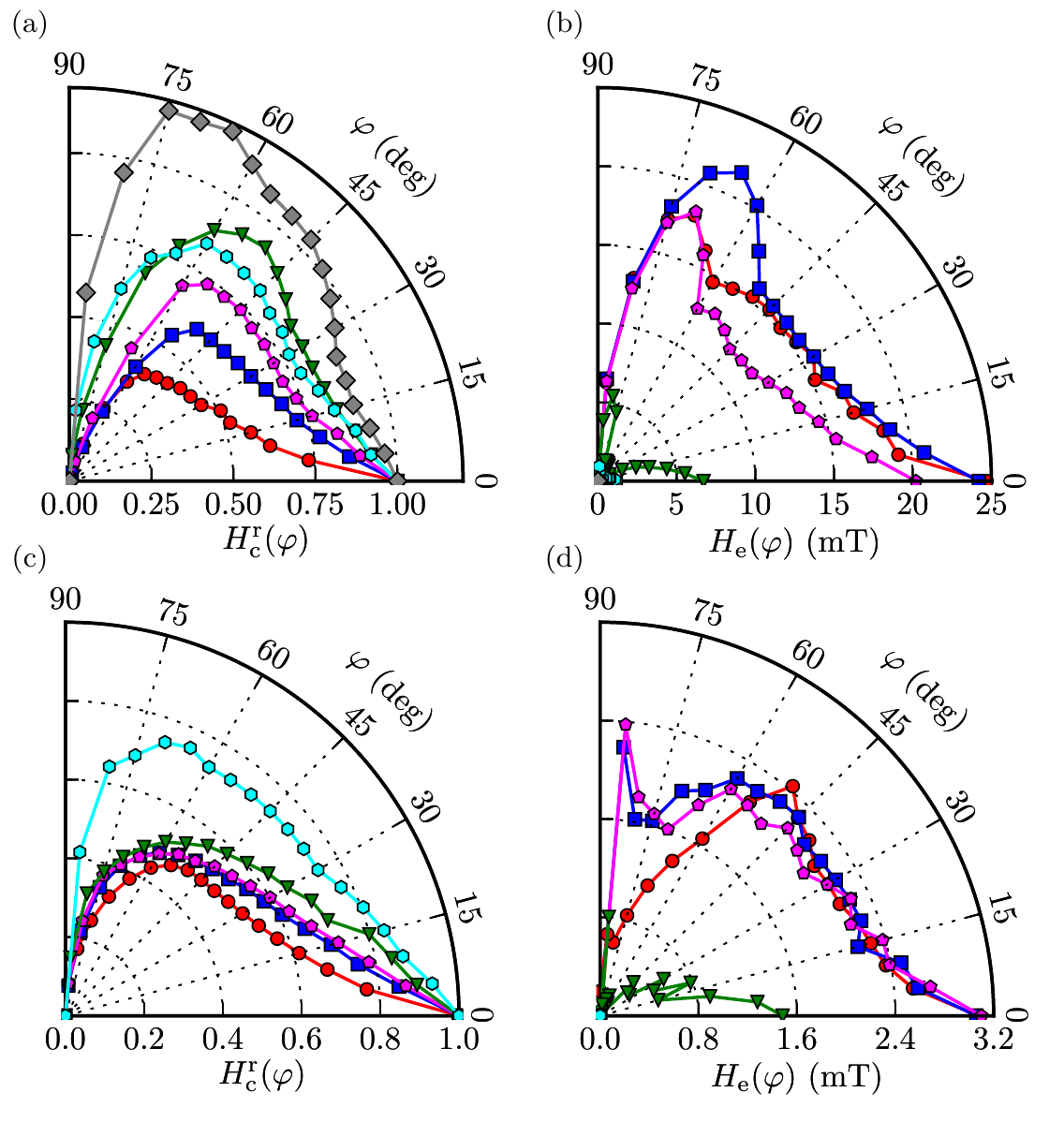}
        \caption{
            \label{fig:S_FMAF_az}
            (a) \Hc  and (b) \He angular thermal evolution  of size-distributed core-shell particles with the (\Kaf,\Jeb) set of parameters.
            (c) \Hc and (d) \He for the second set  $(\Kaf^*,\Jeb^*)$.
            Temperatures are $T = $ \SI{0.01}{K}~(\rond), \SI{10}{K}~(\carre),
            \SI{20}{K}~(\pentagon), \SI{50}{K}~(\trianglemark), \SI{100}{K}~(\pentagondown) and \SI{200}{K}~(\carrebis).
        }
    \end{figure*}

    To create a particle size distribution, a F-AF core-shell particle size distribution is introduced through a F core lognormal distribution as described previously (see Fig.~\ref{domains_twinDM}), with a constant AF shell thickness \taf.
    The nanoparticles are assumed to be all aligned along their easy axis.

    The cycles obtained for different $T$ are shown in Fig.~\ref{fig:S_FMAF_Hevol}(a) (inset), along with the temperature evolution of \Hc (red circles, right axis) and \He (blue squares, left axis).

    The introduction of the F-AF exchange interaction in a size-distributed particle assembly results in an increase of the critical temperature at which \Hc is negligible, as shown in Fig.~\ref{fig:S_FMAF_Hevol}(a).
    This coercive enhancement is an experimental phenomena previously observed in F-AF core-shell nanoparticles.\cite{Skumryev-2003-ID347, Evans-2009-ID352}
    Here, this enhancement is clearly reproduced by the simulation for an assembly of F-AF particles and is due to the additional anisotropy of the rotating AF spins.

    It should be noted that the F-AF exchange interaction results in a non-monotoneous evolution of \Hc with $T$.
    Indeed, a bump is present around \SI{50}{K} as shown in Fig.~\ref{fig:S_FMAF_Hevol}(a).
    The \Hc evolution is the result of the competition between the added AF anisotropy, and the thermal energy.
    A non-monotoneous evolution of \Hc has been previously experimentally observed in ferromagnetic-only nanoparticles.\cite{Zysler-2003-ID442, Pianciola-2015-ID443}
    Such behavior has been attributed to a frustrated magnetic surface state of the particles.\cite{DeBiasi-2006-ID316}
    In our case, the surface state of the particles is an interfacial exchange interaction with an AF.

    Additionally, a decrease of \He with $T$ is observed.
    Indeed, small particles exhibit a low exchange-blocking temperature, as this temperature depends on the anisotropy energy (related to the AF volume) and the interfacial exchange coupling energy (related to the F surface).
    Thus, the AF shell of small MNPs rotate with the F core, and induce an \Hc enhancement.
    On the contrary, the AF shell of bigger MNPs stays frozen, and they do not participate in the \Hc enhancement.
    Instead, they contribute to \He.
    As the temperature increases, the blocked MNPs population decreases, and the exchange bias disappears.

    The reversal cycles (inset in Fig.~\ref{fig:S_FMAF_Hevol}(a)) exhibit an asymmetry of magnetization reversal which evolves with $T$.
    This asymmetry is a specific property of exchange-biased systems, as shown in previous studies.\cite{Spenato-2003-ID482}
    Here, it is demonstrated that this asymmetry is also temperature dependent.

    It is expected that \Hc and \He dependence on $T$ evolves as a function of the F-AF properties (\taf, \Kf, \Kaf, \Jeb, etc.), even for particle systems with the same size distribution.
    To illustrate such a dependence, a second set of parameters ($\Jeb^*$ and $\Kaf^*$) was used, defined in Tab.~\ref{tab:parameters}.
    \Hc and \He dependence on $T$ with this second set of parameters is shown in Fig.~\ref{fig:S_FMAF_Hevol}(b).
	A \He saturation is observed below $T = \SI{40}{K}$, and then $\He$ falls down to zero above.
    This \He saturation is explained by the lognormal kind of distribution, associated with a relatively small ($\Jeb^*, \Kaf^*$) compared to ($\Jeb, \Kaf$).
    Such a phenomenon was previously observed in an experimental study involving diameter size-distributed Fe nanoparticles embedded in a $\mathrm{Cr_2O_3}$ AF matrix.\cite{Sort-2004-ID353}
    The \He maximum value is smaller than for the previous set of parameters (see Tab.~\ref{tab:parameters}), as expected due to $\Jeb^* < \Jeb$.

    \Hc decreases monotoneously.
    It should be noted that the coercive enhancement (relative to F-only size-distributed MNPs) observed in Fig.~\ref{fig:S_FMAF_Hevol}(a) is no longer present.
    Moreover, the strong asymmetry observed in Fig.~\ref{fig:S_FMAF_Hevol}(a) (inset) is not visible anymore.
    Indeed, $\Kaf^*$ is much smaller than \Kaf, as shown in Tab.~\ref{tab:parameters}.

    Finally, the bump is observed for both sets of parameters: a bump of \Hc is observed when \He decreases, at $T = \SI{40}{K}$ in Fig.~\ref{fig:S_FMAF_Hevol}(a) and \SI{60}{K} in Fig.~\ref{fig:S_FMAF_Hevol}(b).
    The amplitude and position of the bump are non-triavial, as these depend on the parameters of the system.
    Indeed, they are related to the apparition rate of the MNPs population with unblocked AF.

	\He and \Hc angular dependence values for both sets of parameters are shown in Fig.~\ref{fig:S_FMAF_az}.
	In both cases, shapes of \Hcr angular dependence with $T$ are similar to the one observed previously in this paper.
    There are still apexes present in the figure, of which the angular position and amplitude evolve with $T$ in a more complex manner.
    In particular, \Hcr values are greater than one in between $\varphi =$ \SI{45}{\degree} and \SI{80}{\degree} at \SI{200}{K} for the (\Kaf, \Jeb) set of parameters, as shown in Fig.~\ref{fig:S_FMAF_az}(a).
    Thus, this phenomena is only present for temperature close to the blocking temperature and for exchange biased MNPs.
    Also , it is not present for the other set of parameters and consequently depends on the ratio between the exchange bias energy and the AF anisotropy.
    It demonstrates that the complex summation of magnetization reversals within an assembly of particles leads to this remarkable phenomena of a \Hcr value greater than one.
    The model presented here reveals such a temperature dependent phenomena as it considers the magnetization reversal of each particle (including superparamagnetic and hysteretic behaviors).

    The \He angular dependence is strongly dependent on the (\Kaf, \Jeb) set of parameters, and also depends on $T$.

	Consequently, probing the thermal dependence of \He and \Hc angular behaviors would contribute to a precise determination of magnetic constants in size-distributed core-shell nanoparticles.

\section{Conclusion}

    In this paper, theoretical modeling of size-distributed F-AF core-shell nanoparticles is proposed, using F and AF uniaxial anisotropy.
    Thermal fluctuations were included with the Néel relaxation theory in a one-valley energy state approximation, by integrating over all the accessible states around the equilibrium valley.
    Also, for both sets of ferromagnetic-only particles and core-shell particles, a particle size distribution was introduced, assuming no inter-particle interactions, in order to reproduce actual systems.

    This process made it possible to calculate temperature-dependent reversal behavior at different external field angles, for both F-only and F-AF exchange-coupled assembly of particles.
    The results obtained for a single F particle are in good agreement with previous models.
    In particular, ferromagnetic to superparamagnetic transition is obtained via multiple state integration.
    The exchange-coupled size-distributed particles show an additional anisotropy carried by the AF, increasing the overall blocking temperature.
    Increase of stability and coercive fields are important for applications such as magnetic storage or magnetic hyperthermia.
    Furthermore, this study shows that size-distributed exchange-coupled nanoparticles exhibit non-monotoneous evolution of \He and \Hc with $T$.
    The angular dependence of \Hc exhibits an apex, whose position and amplitude are strongly $T$-dependent.
    For exchange biased MNPs, it is shown that a reduced coercive field greater than one can be obtained.
    The angular dependence of \He with $T$ exhibits complex behaviors.
    Studying the thermal dependence of \He and \Hc angular behaviors would contribute to a precise determination of magnetic constants in size-distributed core-shell nanoparticles.
    These phenomena are beyond the scope of a non-temperature dependent model, such as MB model.
    Whereas our model does include a size distribution, the development of models including inter-particle interactions and incoherent reversal modes in conjunction with a size-distributed assembly of nanoparticles would be of interest to progress beyond the model presented here.

\begin{acknowledgments}
    This research was supported by the project 29785ZA in the programme Partenariat Hubert-Curien (PHC) and the France/SA NRF Protea grant (grant number 85059).

    The program used for simulation is open-source software written in Python\cite{Oliphant-2007-ID455}, using Scipy\cite{Walt-2011-ID457} and Matplotlib\cite{Hunter-2007-ID454} libraries.
    The source code is freely available under the GPLv3 license on \url{https://github.com/LabMagUBO/StoneX}.
    In addition to the terms of the GPLv3, we kindly request that any work using this program refers to the latter website and this paper.
\end{acknowledgments}

\bibliography{bibliography}

\end{document}